\documentclass[11pt]{article}
\usepackage[centertags]{amsmath}
\usepackage{amsfonts} \usepackage{amssymb} \usepackage{amsthm}
\usepackage{epsfig}

\def\one{{\rm 1\kern -.9mm l}}                             %

\def\beq{\begin{equation}}
\def\eeq{\end{equation}}
\def\beqa{\begin{eqnarray}}
\def\eeqa{\end{eqnarray}}

\newcommand{\bea}{\begin{eqnarray}}
\newcommand{\ena}{\end{eqnarray}}

\usepackage{graphicx}
\usepackage{dcolumn}
\begin{document}

\begin{titlepage}
\vspace*{2cm}
\centerline{\LARGE \bf 
${\cal N}=1$ Super QCD and Fractional Branes}\vskip .5cm
 
\vskip 1.4cm \centerline{\bf Franco Pezzella } 
\vskip .8cm \centerline{\sl
 INFN, Sezione di
Napoli and
 Dipartimento di
Scienze Fisiche, Universit\`a di Napoli } \centerline{\sl Complesso Universitario Monte
S. Angelo, ed. G  - via Cintia -  I-80126 Napoli, Italy} 
 \vskip
2cm

\begin{abstract}
We show how to get the one-loop beta function and the 
chiral anomaly of ${\cal N}=1$ Super QCD from a stack of fractional $N$ D3-branes localized 
inside the world-volume of $2M$ fractional D7-branes on the orbifold $C^3/(Z_2 \times Z_2)$. They are obtained by analyzing the classical supergravity background generated by such a brane configuration, in the spirit of the gauge/gravity
correspondence.
\end{abstract}
\vfill  {\small Talk given at
the IXth International Symposium on Particles, Strings and Cosmology  {\em PASCOS~'03},
 Mumbai-India, January 3-8 2003.} 

\end{titlepage}
\newpage

%

\section{Introduction}

This talk is based on the results of the paper \cite{MNPPS}, that goes in the direction of achieving a deeper understanding of the so-called {\em gauge/gravity correspondence}, originating from the {\em complementarity} between two different but equivalent descriptions of the low-energy properties of D-branes. Indeed, on the one hand a D-brane can be described in terms of the gauge theory living on its world-volume; on the other hand
it is a classical solution of the ten-dimensional supergravity (low-energy string effective theory). Hence one can exploit the classical geometrical properties of $D$-branes to get insight in the {\em dual} gauge theory and viceversa.
This correspondence has led Maldacena to his famous 
conjecture \cite{M}, that provides an exact duality between a conformal and 
highly supersymmetric gauge theory and a (super)gravity theory. It is interesting to apply the gauge/gravity correspondence to less 
supersymmetric and non-conformal gauge theories by using fractional branes
\cite{BDM} which live in an orbifold space, stuck at the orbifold fixed point.
We consider a bound state of $N$ coincident fractional D3-branes and 
$2M$ fractional D7-branes on the orbifold $R^{1,3} \times C^{3}/(Z_2 \times
Z_2)$ yielding ${\cal N}=1$ Super QCD with $M$ fundamental hypermultiplet \cite{BDFM}. The beta function and the chiral anomaly of this theory are reproduced by analyzing
the asymptotic behaviour of the classical supergravity solution \cite{BDFLM2}. 
This is valid in the region
far from the brane, where the ultraviolet properties of the gauge theory 
are reproduced.

\section{Fractional D$p$-branes on the orbifold $R^{1,3} \times 
C^3 / (Z_2 \times Z_2) $ }

We consider fractional D3 and D7 branes on the orbifold
$C^3/(Z_2 \times Z_2)$ in order to study the properties of 
${\cal N}=1$ supersymmetric gauge theories. The orbifold group acts on the directions $x^4, \cdots,
x^9$ transverse to the world-volume of the stack of the $N$ D3-branes. 
The $Z_2 $ group is characterized by two elements
$\{ 1, h \}$,  with $h^2=1$, hence the four elements of the
tensor product ${Z}_2\times {Z}_2$ are easily obtained. The non-trivial elements act on the complex vector $\vec{z}=(z^1=x^4+ix^5,z^2=x^6+ix^7,z^3=x^8+ix^9)$
of $C^3$ as:
$h_1=h~\times~1  \Rightarrow  z_1 \rightarrow z_1, 
z_2 \rightarrow -z_2, z_3 \rightarrow -z_3$; 
$ h_2=1~\times~h \Rightarrow  z_1 \rightarrow -z_1, z_2 \rightarrow z_2, z_3 \rightarrow -z_3$; 
$h_3=h~\times~h  \Rightarrow z_1 \rightarrow -z_1, z_2 \rightarrow -z_2, 
z_3 \rightarrow z_3$.
This orbifold is non-compact with the fixed point $z_1=z_2=z_3=0$,  
corresponding to the three $C^2$ points $z_1,z_2=0$, $z_1,z_3=0$ and $z_2,z_3=0$ 
each of them associated to a vanishing two-cycle ${\cal C}_{2}^{(i)}$ 
with $i=1,2,3$. 
The low energy Type IIB superstring theory spectrum
consists of an {\em untwisted} and three {\em twisted} sectors 
generated by its $p$-form fields dimensionally reduced on the three vanishing two-cycles 
 ${\cal C}_{2}^{(i)}$  ($i=1,2,3$). A fractional D$p$-brane can be regarded
as a D$(p+2)$-brane wrapped on the vanishing two-cycles and, being stuck
at the orbifold fixed point, it couples to all twisted states. The 
Chan-Paton factor accompanying any state of the open string string stretched
between two fractional branes
transforms according to an irreducible (one-dimensional) representation of the orbifold group. 
This means that we have four different kinds of fractional D$p$-branes on this orbifold.
We can get a pure ${\cal N}=1$ $D=4$ SYM with gauge group $SU(N)$ 
by means of a stack of $N$ fractional D3-branes of the same kind. In order to get matter in the fundamental representation
we consider a bound state of $N$ fractional D3-branes and $2M$ fractional
D7-branes of two different kinds, in equal number. Chiral 
matter is
then provided by the open strings having one end attached to the D3-brane and the other one to the D7-brane. Physical states associated to these strings transform under the fundamental representation of the gauge group, while the number of D7-branes can be regarded as a flavor index. We have considered configurations in which the D7-branes extend in the
directions $x^{0}, \dots, x^{3}, x^{6}, \dots, x^{9},$ i.e.
partially along the orbifold, while the D3-branes are in the
directions $x^{0}, \dots x^{3}$, completely localized in the
D7-brane world-volume. In this way we are able to study ${\cal N}=1$ super QCD
with $M$ hypermultiplets. We have shown in \cite{MNPPS} that choosing 
$2M$ D7-branes yields a consistent gauge theory free from 
gauge anomalies.

\section{Supergravity analysis of the dual gauge theory}

The two complementary ways of describing a D$p$-brane 
can be related 
by studying its low-energy effective world-volume action in the supergravity 
background. If a fractional D$p$-brane is thought as a D$(p+2)$-brane wrapped on
the orbifold vanishing two-cycles, 
its world volume action reads:
\begin{eqnarray}
S_{D(p)}& = -\tau_{p+2} \int d^{p+3} \xi e^{\frac{p-1}{4}\phi} \sqrt{ - 
\mbox{det} \left[ G_{\alpha \beta} + e^{ - \frac{\phi}{2}} 
\left(B_{\alpha \beta} + 2 \pi \alpha' {\cal F}_{\alpha \beta} \right) \right]}
 \nonumber \\
{}&+ \tau_{p+2} \int \left( C \wedge e^{B+2\pi \alpha' {\cal F} } \right)_{p+3}
\end{eqnarray}
where $\tau_p= \frac{ (g_s \sqrt{\alpha'})^{-\frac{1}{2}}}{(2 \pi \sqrt{\alpha'} )^{p} }$. 
Specializing this action for $p=3$ and expanding it up to quadratic terms in
${\cal F}$ one has:
\[
S_{D3} = - \frac{1}{g^{2}_{YM}} \int d^4 x \frac{1}{4}{\cal F}_{\alpha \beta} 
{\cal F}^{\alpha \beta} + \frac{\theta_{YM}}{32 \pi^2} \int d^4 x 
{\cal F}_{\alpha \beta} \tilde{\cal{F}}^{\alpha \beta} ,
\]
where
\begin{equation}
\! \! \! \! \!
\frac{1}{ g^{2}_{YM} } \equiv \frac{1}{16 \pi^3 g_s \alpha^{'}} 
\int_{\cup {\cal C}_{2}^{(i)}}
e^{- \phi} B_2 ; \,\,\,
\theta_{YM} \equiv \tau_5 4 \pi^4 { \alpha^{'}}^2 
\int_{\cup {\cal C}_{2}^{(i)}} (C_2 + C_0 B_2). \label{t}
\end{equation}
These equations show the fundamental role played by the 
the twisted fields
$B_{2} = \sum_{i=1}^{3} b^{i}
\omega^{(i)}_{2}$ and $ C_{2} = \sum_{i=1}^{3} c^{i} \omega_{2}^{(i)}$,
being $\omega_{2}^{(i)}$ the volume form dual to the vanishing two-cycle ${\cal C}_{2}^{(i)}$.
The next step is to plug into eqs. (\ref{t}) the supergravity 
background generated by the D3/D7 system. We have determined only 
the asymptotic
behaviour for large distances of the classical solution \cite{MNPPS} 
and this has been revealed sufficient for computing the gauge coupling 
constant and the $\theta$ angle of ${\cal N}=1$ super QCD 
$(z_i = \rho_i e^{i \theta_i})$:
\[
\frac{1}{g^{2}_{YM}} = \frac{1}{16 \pi g_s} + \frac{1}{8 \pi^2} \left(
N \sum_{i=1}^{3} \mbox{log} \frac{\rho_i}{\epsilon} - M \mbox{log} \frac{\rho_1}{\epsilon} \right), \,\,\,
\theta_{YM} = -N \sum_{i=1}^{3} \theta_{i} + M \theta_1 .
\] 
The coordinates $z_i=\rho_i e^{i\theta_{i}}$ are holographically identified
with the scalar components  $\phi_i$ of the superfield $\Phi_i$ appearing
in the cubic superpotential $W=\mbox{Tr} \left( \Phi_1 \left[ \Phi_2, \Phi_3
\right] \right)$ \cite{BDFM}. The scale transformation $\phi_i \rightarrow \mu \phi_i$
and the $U(1)_R$ one $\phi_i
\rightarrow 
e^{i \frac{2}{3} \alpha} \phi_i$ induce on $z_i$ the transformation $z_i \rightarrow \mu e^{i 2 \alpha/3} z_i$, i.e.  
$\rho_i \rightarrow \mu \rho_i$ and $\theta_i \rightarrow \theta_i + 2/3 
\alpha$. These latter act on the gauge parameters as follows:
\begin{equation}
\frac{1}{g^2_{YM}} \rightarrow \frac{1}{g^2_{YM}} + \frac{3N-M}{8 \pi^2}
\mbox{log} \mu, \,\,\, \, \, \theta_{YM} \rightarrow \theta_{YM} - 2\alpha 
\left( N - \frac{M}{3} \right).
\end{equation}
The first equation generates the one-loop $\beta$-function of ${\cal N}=1$ Super QCD with
$M$ hypermultiplets:
\[
\beta(g_{YM}) = - \frac{3N-M}{8 \pi^2} g^{3}_{YM},
\]
while the second one reproduces the chiral $U(1)$ anomaly.

{\bf Acknowledgemnts.} I would like to thank P. Di Vecchia, A. Liccardo, R. Marotta, F. Nicodemi, R. Pettorino and F. Sannino for useful discussions.

\end{document}